\documentclass[12pt]{article}
\usepackage{amssymb}
\usepackage{amsmath}
\usepackage{amsfonts}
\usepackage{color}
\usepackage{cite}
\usepackage{amsmath, amsfonts, amsthm, amssymb, graphicx}
\usepackage{epstopdf}
 \usepackage{slashed}
 \usepackage{graphicx,epsfig}
\usepackage{amsmath}
\usepackage {amssymb}
\usepackage[utf8]{inputenc}
\usepackage{hyperref}
\usepackage[T1]{fontenc}

\definecolor{blue-violet}{rgb}{0.54, 0.17, 0.89}
\definecolor{PineGreen}{cmyk}{0.92, 0, 0.59, 0.25}
\definecolor{OliveGreen}{cmyk}{0.64, 0, 0.95, 0.40}
\definecolor{RawSienna}{cmyk}{0, 0.72, 1, 0.45}
\definecolor{Gray}{cmyk}{0, 0, 0, 0.50}
\definecolor{MidnightBlue}{cmyk}{0.98, 0.13, 0, 0.43}
\definecolor{Orange}{cmyk}{0, 0.61, 0.87, 0}
\definecolor{LimeGreen}{cmyk}{0.50, 0, 1, 0}
\definecolor{Green}{cmyk}{1, 0, 1, 0}

\usepackage[makeroom]{cancel}
\usepackage[normalem]{ulem}

\numberwithin{equation}{section}

\begin{document}

\title{\bf Naked BPS singularities in AdS$_{3}$ supergravity}
\author{Gaston Giribet$^{1,}$\thanks{{gaston.giribet@nyu.edu}}\;,  Olivera Mi\v{s}kovi\'{c}$^{2,}$\thanks{olivera.miskovic@pucv.cl}\;, Nahuel Yazbek$^{3,}$\thanks{nyazbek@dm.uba.ar}\;, Jorge Zanelli$^{4,5}$\thanks{jorge.zanelli@uss.cl, z@cecs.cl (corresponding author)}
\bigskip\\
{\small $^1$  Department of Physics, New York University,}\\ {\small {726 Broadway, New York City, 10003, USA.}}\smallskip\smallskip\\
{\small $^2$ Instituto de F\'\i sica, Pontificia Universidad Cat\'olica de Valpara\'\i so,}\\ {\small Avda.~Universidad 330, Curauma, Valpara\'{\i}so, Chile.}\smallskip\smallskip\\
{\small $^3$ Instituto de Matem\'atica Luis A. Santal\'o, CONICET-UBA,}\\ {\small Ciudad Universitaria, Pabell\'on I, CABA, C1428ZAA, Argentina.}\smallskip\smallskip\\
{\small $^4$ Centro de Estudios Cient\'\i ficos (CECs), Arturo Prat 514, Valdivia, Chile.}\\
{\small $^5$ Universidad San Sebastián, General Lagos 1163, Valdivia, Chile.}
}

\date{}

\maketitle

\begin{abstract}
AdS supergravity admits supersymmetric solutions that describe BPS defects. We investigate such solutions in AdS$_3$ supergravity formulated as a Chern-Simons theory on $\mathrm{OSp}(2|1)\,\times\, \mathrm{OSp}(2|1)$ and compute the  Killing spinor equation on the BTZ geometry, looking for BPS solutions on the entire space of parameters. We focus our attention on defects that represent geometries with integer angular excesses corresponding to specific negative values of the BTZ mass, extending other results in the literature to all real values of mass and angular momentum. We argue that, in the semiclassical limit, the BPS defects can be associated with degenerate representations of the Virasoro symmetry at the boundary. The case of non-diagonal representations, describing stationary, non-static defects, is also discussed.   
\end{abstract}

%\tableofcontents
\bigskip

%%%%%%%%%%%%%%%%%%%%%%%%%%%%
\section{Introduction} % 1 %
%%%%%%%%%%%%%%%%%%%%%%%%%%%%
Stanley Deser was a pioneer in fields of supergravity \cite{Deser-Zumino} and $2+1$ gravity \cite{DJtH}. It is therefore fitting that we dedicate this article on supersymmetric states in $2+1$ gravity to him, as he was an inspiration to all of us working on his footsteps.

The BTZ geometry \cite{BTZ} is a solution to Einstein equations with negative cosmological constant in $2+1$ dimension, which, for a certain range of its two parameters, describes an asymptotically anti-de Sitter (AdS) \cite{BH} black hole. Being a three-dimensional Einstein space, the BTZ solution is locally equivalent to AdS$_3$ spacetime itself, which means that the former can be constructed by identification from the latter \cite{BHTZ}. In particular, this implies that locally, the solution is of constant negative curvature, albeit with a curvature $\delta$-like singularity at center \cite{Olivera-Z,Briceno-Martinez-Z}. 

Similar to the four-dimensional Kerr solution, the BTZ metric has two integration constants, $M$ and $J$, which are the Noether charges associated with the two Killing vectors that generate the $\mathbb{R}\times SO(2)$ isometry. Consequently, they are interpreted as the mass and the angular momentum, respectively. For the solution to represent a black hole, the constraints $M\geq |J|/\ell\geq 0$ must hold, with $\ell$ being the curvature radius of AdS$_3$ space; extremal black holes correspond to $M= \pm J/\ell> 0$. Stationary BTZ black holes share many geometrical properties with their higher-dimensional analogs: they exhibit an event horizon and an inner Killing horizon, along with an ergosphere and a shielded singularity. It also shares with higher-dimensional black holes their main thermodynamics properties, such as finite Hawking temperature and a Bekenstein-Hawking entropy that obeys the area law. 

As noted in \cite{Witten}, the existence of these black holes with non-trivial thermodynamic properties makes three-dimensional Einstein gravity a much more exciting system, especially when studied in connection with the AdS/CFT correspondence \cite{Strominger}. In addition, the BTZ black hole appears in many other scenarios: being locally equivalent to AdS$_3$, it is also a solution to supergravity \cite{Coussaert:1993jp}, conformal gravity \cite{CG}, other Chern-Simons actions \cite{Garcia}, string theory \cite{HorowitzWelch}, topologically massive gravity \cite{TMG}, higher-curvature gravity \cite{TMG, NMG}, bi-gravity theory \cite{Banados}, and higher-spin theories \cite{Krauss}; see also  \cite{1106.4788,1108.2567,1203.0015,1302.0816,1309.4362,1402.1465,1301.0847,1207.2844,1512.00073,1111.3381}.
For nonsingular asymptotically BTZ solutions, see, for instance,  \cite{Edery:2020kof,Edery:2022crs}.

The BTZ geometry describes a stationary black hole only for $M\geq |J|/\ell\geq 0$, but it also presents interesting features in other ranges of parameters. For example, in the case $M+1/(8G)=J/\ell =0$, with $G$ the Newton constant, BTZ reduces to global AdS$_3$ spacetime. In the segment $0>M>-1/(8G)$ ({and $M<-|J|/\ell$}), the solution can be regarded as point-like massive particles, i.e., a naked conical singularity similar to the point-like particles of three-dimensional gravity in flat space \cite{DJtH}. Less studied cases correspond to the range 
$M \ell \leq -|J|$, $(M\ell)^2-J^2>1/(8G)^{2}$, where the solution describes geometries with angular excesses around the origin. Here, we will be concerned with the latter; we will consider solutions for the specific negative values of $M+J/\ell$, such that they exhibit integer angular excesses and special supersymmetric properties. From the AdS/CFT point of view, in the semiclassical limit such geometries are associated with special non-normalizable states in the dual CFT$_2$. Static supersymmetric configurations with negative $M$ correspond to degenerate representations in the dual CFT$_2$, while stationary non-static configurations with {$M\pm J/\ell<0$} can be identified with non-diagonal (spinfull) non-integrable representations of the type studied by Migliaccio and Ribault in \cite{Migliaccio}.

Solutions that represent {naked singularities} of integer angular excesses have been recently considered in the literature, especially in connection to higher-spin theories. In the higher-spin context they were introduced in \cite{1111.3381}, where the way to characterize such defects as states with trivial Chern-Simons holonomy was explained in detail. A clear discussion on the holographic interpretation of those states appeared in \cite{1210.8452}, and the identification between conical solutions and primaries in the $W_s$ minimal models that appear in spin-$s$ theories was revisited in \cite{1712.08078}. The role of integer angular excess solutions for the $s=2$ case was studied in \cite{1412.0278} and more recently in \cite{2012.07934}; see also references thereof. Other interesting papers where the defects and degenerate representations are discussed in the context of three-dimensional gravity are \cite{1811.05710, 2004.14428}. There are also interesting works in two dimensions that, to some extent, are related to this, e.g. \cite{1904.05228, 2011.01953, 2305.19438}; however, the relation to the uplift to dimension three is not obvious to us. Here, we will be concerned with naked singularities of arbitrary values of the parameters in three-dimensional supergravity. We will show that those geometries that exhibit integer angular excesses are the only BPS states appearing in the negative mass sector of the BTZ geometry \cite{Olivera-Z}. It is well-known that the positive mass sector admits supersymmetric solutions: the massless BTZ solution $M=J=0$ has two exact supersymmetries, while the extremal solutions with $M=\pm J/\ell\neq 0$ have only one \cite{Coussaert:1993jp}. These, together with the AdS$_3$ vacuum, are the only solutions of positive mass with supersymmetry. 

Here we want to investigate the BPS solutions in the negative mass sector: we will solve the spinor Killing equation in the stationary geometry and look for globally well-defined solutions that preserve at least one supersymmetry. The latter would represent BPS solutions with naked singularities. BPS solutions with naked singularities in AdS supergravity are known to exist in other models, e.g., the Romans solution of $\mathcal{N}=2$ supergravity in AdS$_4$ is $\frac 12$ BPS \cite{Romans}. Here, we will identify similar solutions in AdS$_3$. We will study the case of Chern-Simons theory for the $\mathrm{OSp}(2|1)\times \mathrm{OSp}(2|1)$ supergroup, which realizes a supersymmetric extension of AdS$_{3}$ algebra with $\mathcal{N}=2$ supercharges \cite{Achucarro:1986uwr}.\footnote{Three-dimensional Einstein gravity in AdS space is equivalent, at the level of actions, to Chern-Simons gravity for AdS$_{3}\simeq \mathrm{SO(2,2)}$. This equivalence includes boundary terms, where Chern-Simons action provides correct surface terms which make AdS gravity finite \cite{Miskovic:2006tm}.} Negative mass BPS states were also studied in \cite{Izquierdo:1994jz} in the context of (2,0) AdS$_3$ supergravity, in agreement with our results can be found. Similar solutions were also studied in the context of Kaluza-Klein compactified type IIB superstrings \cite{Balasubramanian:2000rt}. Such BPS defects, on the other hand, can be seen to persist in the flat space limit \cite{Donnay}. As we will argue, in the AdS$_3$ case these defects are associated with degenerate representations in the dual CFT$_2$, which in the semiclassical limit of three-dimensional gravity can be associated with an effective Liouville field theory \cite{vanDriel}; see also \cite{Krasnov, Donnay2}. A holographic interpretation of these conical defects as degenerate representations of the chiral algebra in the context of higher spin theory was also discussed in \cite{Hikida:2012}.

%%%%%%%%%%%%%%%%%%%%%%%%%%%%%%%%%%%%%
\section{BPS BTZ geometries} % 2 %
%%%%%%%%%%%%%%%%%%%%%%%%%%%%%%%%%%%%%%
\subsection{Stationary metrics} % 2.1%
%%%%%%%%%%%%%%%%%%%%%%%%%%%%%%%%%%%%%%

The BTZ geometries are solutions of Einstein's equations in 2+1 dimensions with negative cosmological constant described by the metric
\begin{eqnarray}
ds^{2} &=&-f^{2}dt^{2}+\frac{dr^{2}}{f^{2}}+r^{2}\left( Ndt+d\varphi \right)
^{2},  \notag \\
f^{2} &=&-M+\frac{r^{2}}{\ell ^{2}}+\frac{J^{2}}{4r^{2}}\,,\qquad N=-\frac{J%
}{2r^{2}}\,,  \label{BTZ metric}
\end{eqnarray}%
where $t\in \mathbb{R}$, $r\in \mathbb{R}_{\geq 0}$ and $\varphi \in [0,2\pi ]$, with $\ell $ being the curvature radius of the AdS space (hereafter we will take Newton's constant $G=1/8$ unless otherwise stated). As it is well-known \cite{BHTZ}, depending on the parameters $(M,J)$, the geometries \eqref{BTZ metric} correspond to: 
\begin{equation*}
\fbox{$\begin{array}{ll}
M\geq |J|/\ell \geq 0:\medskip  & \text{black hole with mass }M\geq 0%
\text{ and angular momentum }J\text{;} \\
M=-1\text{, }J=0:\medskip  & \text{globally AdS}_{3}\text{ space;} \\
{M\leq -|J|/\ell }\text{{,}}{\ M^{2}-(J/\ell )^{2}<1}:\medskip  & \text{%
naked singularity with angular deficit (point-like particle);} \\
{M\leq -|J|/\ell }\text{{, }}{M^{2}-(J/\ell )^{2}>1}:\medskip  & \text{naked
singularity with angular excess;} \\
|M|<|J|/\ell : & \text{over-spinning geometries.}
\end{array}$}
\end{equation*}

Geometries with zero or negative mass parameter $M\neq -1$ describe topological defects whose total angle in the spatial plane is $2\pi \left( 1-\alpha \right) $. Thus, the parameter $\alpha $ measures a difference with respect to the space without defect, $\alpha =0$, corresponding to the AdS$_3$ space. Furthermore, since the angular deficit is introduced in a plane by Killing vector identifications, to define a true manifold, successive identifications must yield the identity operation after a finite number of iterations. Consequently, the deficit angle, $\alpha$, must be a rational fraction of $2\pi $, i.e. $\alpha/(2\pi)\in\mathbb{Q}$. In the rotating case, two angular deficits turn out to be associated with two rational numbers \cite{Briceno-Martinez-Z}.

The coordinate frame where the static defect becomes explicit is
\begin{equation}
ds^{2}=-\left( \frac{\rho ^{2}}{\ell ^{2}}+1\right) \,d\tau ^{2}+\frac{d\rho
^{2}}{\frac{\rho ^{2}}{\ell ^{2}}+1}+(1-\alpha )^{2}\rho ^{2}d\varphi ^{2}\,,
\end{equation}
where $0\leq \varphi \leq 2\pi $ is periodic. Introducing the coordinates $(t,r)= \left(\tau/(1-\alpha),(1-\alpha)\rho \right)$, the metric acquires the form \eqref{BTZ metric}, 
\begin{equation}
ds^{2}=-\left( \frac{r^{2}}{\ell ^{2}}+(1-\alpha )^{2}\right) dt^{2}+\frac{dr^{2}}{\frac{r^{2}}{\ell ^{2}}+(1-\alpha )^{2}}+r^{2}d\varphi ^{2}\,,
\end{equation}
that enables to identify the mass and angular momentum as $M=-(1-\alpha )^{2}$ and $ J=0$. Note that the angular defects produce a conical singularity at $r=0$ in the $\Sigma_{12}$ plane, $(x^{1},x^{2})=(r\cos \varphi _{12}\,,r\sin \varphi _{12})$, where $\varphi _{12}=2\pi \left( 1-\alpha \right) \varphi $, such that \cite{Olivera-Z}
\begin{equation}
R^{ab}+\frac{1}{\ell ^{2}}\,e^{a}\wedge e^{b}=2\pi \alpha \,\delta (\Sigma
_{12})d\Omega _{12}\,J_{12}\,\eta ^{[12][ab]}\,,\qquad T^{a}=0\,,
\end{equation}
where $e^a=e^a_\mu dx^\mu$ is the dreibein 1-form, $d\Omega_{12}$ is the  volume element of $\Sigma_{12}$, and $J_{12}$ is the rotation generator in the plane $\Sigma_{12}$. The curvature and torsion 2-forms are 
\begin{equation}
R_{a}=\frac{1}{2}\,\varepsilon _{abc}R^{bc}=d\omega
^{a}+\frac{1}{2}\,\varepsilon _{abc}\omega ^{b}\omega ^{c}\,, \qquad T^{a}=De^{a}\,,
\end{equation}%
respectively, and $\eta ^{[ab][cd]}=\eta^{ac}\eta^{bd}-\eta^{ad}\eta^{bc}$ is a Lorentz invariant tensor.

%%%%%%%%%%%%%%%%%%%%%%%%%%%%%%%%%%%%%%%%%%%%%%%
\subsection{Chern-Simons Supergravity} % 2.1 %
%%%%%%%%%%%%%%%%%%%%%%%%%%%%%%%%%%%%%%%%%%%%%%%

Geometries \eqref{BTZ metric} describe purely bosonic solutions of three-dimensional supergravity in (A)dS$_3$, whose action can be expressed as a Chern-Simons (CS) theory of level $k=\ell/(4G)$ for the supersymmetric extension of AdS$_{3}$ algebra with $\mathcal{N}=2
$ supercharges, given by $\mathrm{osp}(2|1)\times \mathrm{osp}(2|1)$
superalgebra\footnote{The notation used in this text is consistent with \cite{Coussaert:1993jp}
and \cite{Giacomini:2006dr}.} \cite{Achucarro:1986uwr}: 
\begin{equation}
\lbrack J_{a}^{\pm },J_{b}^{\pm }]={{\varepsilon_{ab}}^{c}\,J^{\pm }_{c}}\,,\qquad
[J_{a}^{\pm },Q_{\alpha }^{\pm }]=-\frac{1}{2}\,(\Gamma _{a})_{\alpha
}^{\ \beta }\,Q_{\beta }^{\pm }\,,\qquad \{Q_{\alpha }^{\pm },Q_{\beta
}^{\pm }\}=(C\Gamma ^{a})_{\alpha \beta }\,J_{a}^{\pm }\,.
\end{equation}
Here, $a,b,c=0,1,2$ are Lorentz indices; $\alpha, \beta = 1,2$ are spinor indices; and $\pm $ refer to two commuting copies of the superalgebra. The corresponding gauge field ($A=A_{\mu}\,dx^{\mu}$) is expressed in terms of the dreibein ($e^a$), the Lorentz (spin) connection is $\omega^{ab}=\omega^{ab}_{\ \mu}\,dx^{\mu}$, and the algebra generators as follows,
\begin{equation}
A=\left( \omega ^{a}+\frac{1}{\ell }\,e^{a}\right) J_{a}^{+}+\left( \omega
^{a}-\frac{1}{\ell }\,e^{a}\right) J_{a}^{-}+\frac{1}{\sqrt{\ell }}\,\left(
\xi _{+}^{\alpha }Q_{\alpha }^{+}+\xi _{-}^{\alpha }Q_{\alpha }^{-}\right)
\,,
\end{equation}
where $\omega ^{ab}=-\varepsilon ^{abc}\,\omega _{c}$, $\omega _{a}=\frac{1}{2}\,\varepsilon _{abc}\omega ^{bc}$ and $J^{ab}=\varepsilon ^{abc}\,J_{c}$, $J_{a}=-\frac{1}{2}\,\varepsilon _{abc}\,J^{bc}$. Covariant derivatives on Lorentz vectors and spinors are
\begin{equation}
De^{a}=de^{a}+\varepsilon ^{abc}\omega _{b}e_{c}\,,\qquad D\psi =d\psi -
\frac{1}{2}\,\theta  \omega ^{a}\Gamma _{a}\psi \,.  \label{Lcov}
\end{equation}

We use the representation of Gamma matrices
\begin{equation}
\Gamma _{0}=\mathrm{i}\theta \sigma _{1}\,,\qquad \Gamma_{1}=\theta \sigma _{2}\,,\qquad \Gamma _{2}=\theta \sigma _{3}\,, \label{Gamma}
\end{equation}
where $\theta=\pm 1$ corresponds to the two inequivalent representations of the three-dimensional Clifford algebra $\{ \Gamma_a, \Gamma_b \}=2\eta_{ab}$.  The signature of Minkowski metric is $\eta_{ab}=\mathrm{diag}(-,+,+)$, and the convention for the Levi-Civita symbol is $\varepsilon _{012}=-\varepsilon ^{012}=1$.

The spinorial representation of AdS$_3$ generators becomes
\begin{equation}
P_{a}=\frac{1}{2}\,\Gamma _{a}\,,\qquad J_{a}=\frac{1}{2}\,\theta  \,\Gamma
_{a}\,,
\end{equation}
where $J_{ab}=\frac{1}{4}\,[\Gamma _{a},\Gamma_b]=-
\varepsilon_{abc}\,J^c$. In this representation, the AdS$_3$ covariant derivative acts on a spinor as
\begin{eqnarray}
\nabla \psi  =\left( d+\frac{1}{2}\,\omega ^{ab}J_{ab}+\frac{1}{\ell }%
\,e^{a}P_{a}\right) \psi   =D\psi +\frac{1}{2\ell }\,e^{a}\Gamma _{a}\psi \,,  \label{covariant}
\end{eqnarray}%
where the Lorentz-covariant derivative $D$ is given in (\ref{Lcov}). The extra $\theta $ does not affect the vectorial representation of the generators, as in the definition of $De^{a}$.

%%%%%%%%%%%%%%%%%%%%%%%%%%%%%%%%
\subsection{Killing spinors}  % 3 %
%%%%%%%%%%%%%%%%%%%%%%%%%%%%%%%%
For the metric \eqref{BTZ metric}, the vielbein and a torsionless spin connection can be chosen as
\begin{equation}
\begin{array}{llllll}
e^{0} & =f\,dt\,,\qquad  & e^{1} & =\dfrac{dr}{f}\,,\qquad  & e^{2} & =r(d\varphi + N\, dt\,),\medskip  \\ 
\omega ^{0} & = f d\varphi \,, & \omega ^{1} & =  \dfrac{J}{2f r^2}\,dr\,, & \omega ^{2} &= \dfrac{r}{\ell^2}\,dt - \dfrac{J}{2r}\,d\varphi \,.
\end{array}
\end{equation}

A Killing spinor is a globally defined solution of the condition of invariance under supersymmetry of a state where the fermionic fields $\xi$ vanish, namely,
\begin{equation}
\delta_{\psi }\xi = D\psi +\frac{1}{2\ell}\,\Gamma_a\,e^{a}\psi \equiv d\psi -\frac{1}{2}\,\Gamma_a\left(\theta \omega^a - \frac{1}{\ell}\,e^a \right) \psi =0\,. \label{KSeq}
\end{equation}

With the representation \eqref{Gamma}, the Killing spinor equation \eqref{KSeq} now reads 
\begin{equation}
d\psi +\left[ \frac{1}{2\ell }\left( \rule{0pt}{12pt}f\Gamma _{0}-\left(
\frac{J}{2r}+\theta \, \frac{r}{\ell }\right) \Gamma _{2}\right) (dt-\theta \ell d\varphi )-\frac{1}{2rf}\,\Gamma _{1}\left( \frac{\theta J}{2r}-\frac{r}{\ell }\right) dr
\right] \psi =0\,.
\end{equation}

In coordinates $x^{\pm} =t\pm \theta \ell \varphi$, $\partial _{\pm } = \frac{1}{2\ell }\,\left(\ell \partial _{t}\pm \theta \partial_{\varphi }\right)$, this equation takes the form 
\begin{eqnarray}
0 &=&\partial _{+}\psi \quad \Rightarrow \quad \psi =\psi \left(x^{-},r\right) \,,  \notag \\
0 &=&\partial _{-}\psi +\frac{1}{2\ell}\,\left[ f\Gamma _{0}-\Gamma _{2}\left(
\frac{J}{2r}+\theta \,\dfrac{r}{\ell }\right) \right] \psi \,,  \label{set} \\
0 &=&\partial _{r}\psi -\frac{1}{2rf}\,\Gamma _{1}\left( \dfrac{\theta J}{2r}-\frac{r}{\ell }\right) \psi \,.  \notag
\end{eqnarray}

The solution for generic $(J, M)$ reads\footnote{For details, see arXiv:2402.00171[hep-th].}
\begin{equation}
\psi= \mathrm{e}^{-\frac{\mathrm{i}}{2\ell }\omega\,x^{-}} \left( 
\begin{array}{c}
\left(\frac{r}{\ell} +\frac{\theta J}{2r} -\mathrm{i}\omega \right)^{1/2} \\
-\mathrm{i}\theta \left(\frac{r}{\ell} +\frac{\theta J}{2r} +\mathrm{i}\omega \right)^{1/2} 
\end{array}\right)\eta_1 + \mathrm{e}^{\frac{\mathrm{i}}{2\ell}\omega\,x^{-} }\left( 
\begin{array}{c}
\left(\frac{r}{\ell} +\frac{\theta J}{2r} +\mathrm{i}\omega \right)^{1/2} \\
-\mathrm{i}\theta \left(\frac{r}{\ell} +\frac{\theta J}{2r} -\mathrm{i}\omega \right)^{1/2} 
\end{array}\right) \eta_2\,, \label{rotating}
\end{equation}
where $\eta_1$, $\eta_2$ are constants that might be assumed to be Grassmann numbers, and
\begin{equation}
\omega=\sqrt{-M-\frac{\theta J}{\ell }}\,,\qquad \omega^2\geq 0\, \quad\mbox{or}\quad  -M\ell \geq  \theta J\,.
\end{equation}
Here we take $\omega^2>0$  because we are interested in naked singularities; black hole solutions were discussed in \cite{Coussaert:1993jp}. 

This solution is globally well-defined when the spinor is periodic or antiperiodic in the angle $x^-$, leading to the condition
\begin{equation}
\omega=n\in \mathbb{Z}_{\geq 0}\quad \Rightarrow \quad M+\frac{\theta J}{\ell}=-n^{2}\,. \label{global}
\end{equation}
More precisely, the spinor \eqref{rotating} is periodic for even $n$, and anti-periodic for odd $n$.

This means that the BPS states occur along straight lines in the $M$-$J$ plane, as shown in Fig.~1. The Killing spinors \eqref{rotating} contain up to two independent constants of integration, $\eta_1$, $\eta_2$. Hence, for fixed values of $\omega=n$ and $\theta$ corresponding to one dotted line in Fig.~1, Eqs.~\eqref{set} admit a two-dimensional space of solutions labeled by $(\eta_1, \eta_2)$.
The configuration $M=0=J$ has two Killing spinors with periodic boundary conditions, whereas $M=-1$, $J=0$ has two antiperiodic spinors. In general, the geometries at the intersection points $(n,m)$ have two Killing spinors, corresponding to two inequivalent representations of gamma matrices for $\theta=\pm1$, which are periodic or antiperiodic depending on the signs of $(-1)^n$ and $(-1)^m$.

\begin{figure}%[H]
    \centering
\includegraphics[width=1.1\textwidth]{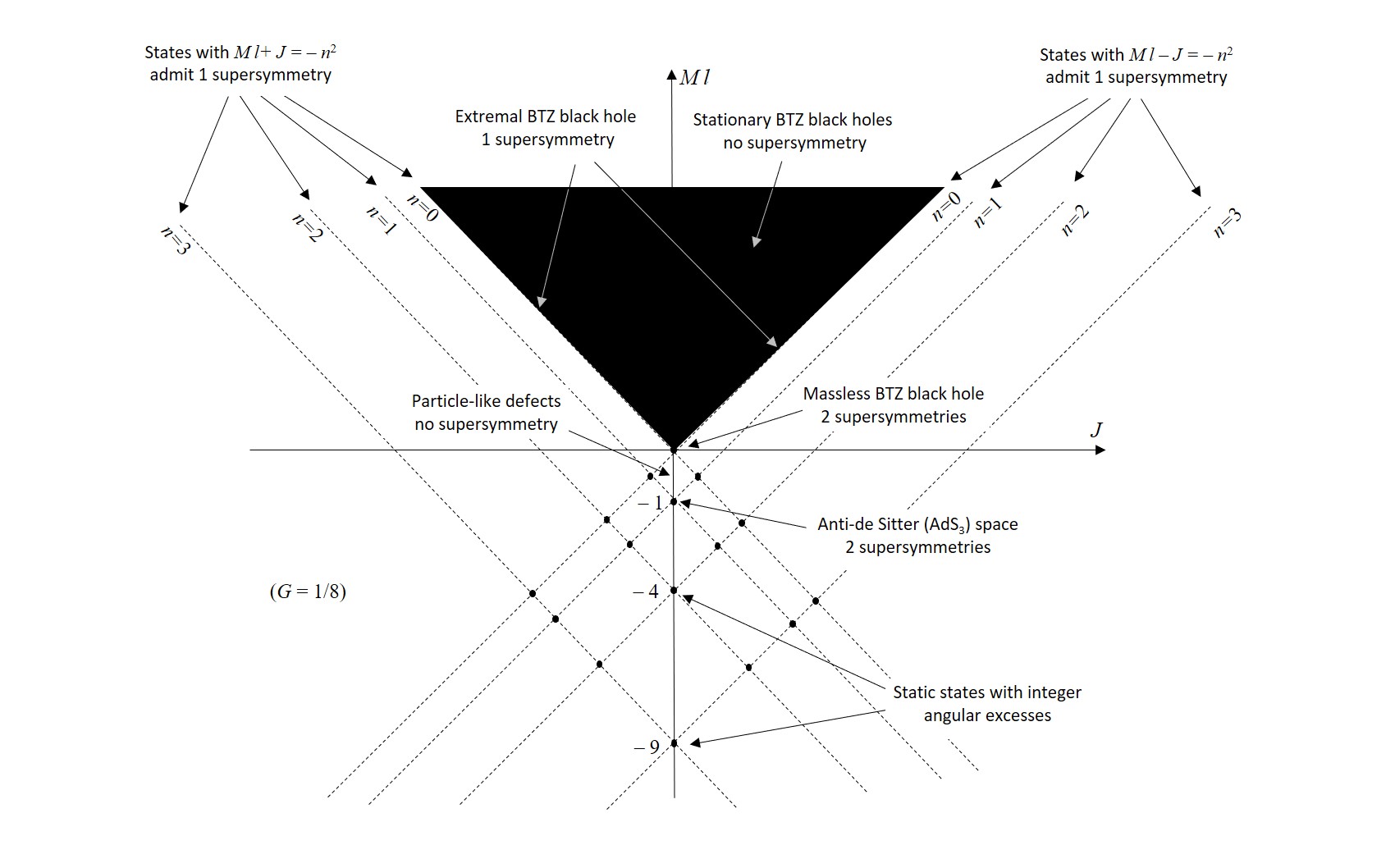}
\caption{BPS states in the ($M\ell ,J$) plane in units $G=1/8$. Half BPS states, including extremal BTZ black holes, correspond to the lines $M\ell \pm J=-n^2$ with $n\in \mathbb{Z}$, having either left-moving ($-$) or right-moving ($+$) KS. States located at the intersection of two such lines admit two KS. In particular, we have AdS$_3$ space among the latter and the $M=J/\ell=0$ BTZ geometry, the latter being the vacuum with periodic boundary conditions.}
\label{BPSstates}
\end{figure}

Not all points in the lines $M\pm J/\ell = n^2$ correspond to genuine manifolds. Only those for which the identification Killing vectors correspond to rotations by rational fractions of $2\pi$ meet this requirement. It is easy to show that those values of mass and angular momentum satisfy $M,\,J\, \in \mathbb{Q}$ \cite{Briceno-Martinez-Z}.

%%%%%%%%%%%%%%%%%%%%%%%%%%%%%%%%%%%%%%%%%%%%%%%%%%%%%%%%%%%%%%%%
\section{A CFT boundary perspective}
%%%%%%%%%%%%%%%%%%%%%%%%%%%%%%%%%%%%%%%%%%%%%%%%%%%%%%%%%%%%%%%%

Before concluding, let us make some comments about how to look at these special states from the holographic point of view; that is to say, from the dual CFT$_2$ perspective. We are interested in identifying which representations of the Virasoro symmetry are those associated with the negative mass BPS configurations discussed above. As anticipated, in the semiclassical limit, the static BPS states can be associated with the so-called degenerate representation of Liouville field theory, which are non-normalizable states of the CFT$_2$ that contain null descendants in the Verma module. But, first, let us review where the Liouville CFT$_2$ description comes from. Since here we are dealing with AdS$_3$ supergravity, the right theory to look at would be super-Liouville; nevertheless, it will be sufficient for us to focus on the bosonic theory first; we will see below how the super-Liouville leads to the same results. 

In the renowned paper \cite{BH}, Brown and Henneaux found that the asymptotic isometries in asymptotically AdS$_3$ spacetimes is generated by two commuting copies of the Witt algebra, with the associated Noether charges satisfying two copies of Virasoro algebra with the central charge\footnote{In this subsection, we restore the dependence of the Newton constant $G$.}
\begin{equation}
c=\frac{3\ell}{2G}\,.\label{cBH}
\end{equation}
The extension of the study of the asymptotic dynamics in AdS$_3$ supergravity was done in \cite{Banados:1998pi} and yields an equivalent result. The observation in \cite{BH} is often considered a precursor of AdS/CFT \cite{Malda}, the reason being that, from a modern perspective, the central charge (\ref{cBH}) is understood as that of the dual CFT$_2$. Also, one can identify the conformal dimension of the CFT$_2$ states as coming from the $L_0$ and $\bar L _0$ generators of the isometry algebra, which correspond to the spin and the energy of the configuration; namely
\begin{equation}
h-\frac{c}{24}=\frac 12 (\ell M+J) \, , \ \ \ \ \ \bar h-\frac{c}{24}=\frac 12 (\ell M-J),
\end{equation}
which is equivalent to 
\begin{equation}
\frac{ h+\bar h }{\ell}=  M+\frac{1}{8G}   \, , \ \ \ \ \ h-\bar h=J.
\end{equation}
This result enables us to identify the gap in the spectrum of the BTZ black hole with respect to the AdS$_3$ vacuum $h=\bar h =0$, which corresponds to
\begin{equation}
M_0=-\frac{1}{8G} \, , \ \ \ \ \ J_0=0 .\label{ffff}
\end{equation}
Reciprocally, the configuration $M=J=0$ corresponds to
\begin{equation}
h_0=\bar{h}_0=\frac{c}{24}.\label{Gfff}
\end{equation}
In \cite{vanDriel}, van Driel, Coussaert and Henneaux went further in the CFT$_2$ description of the AdS$_3$ asymptotic dynamics and, following a sequence of steps that includes a gauge fixing and a prescription of boundary conditions that implement a Hamiltonian reduction \cite{Donnay2}, they found that the asymptotic dynamics of Einstein (super-)gravity in AdS$_3$ is governed by a (super-)Liouville field theory action. {Because this equivalence is valid only at the level of the actions, one should not understand the relation between three-dimensional gravity and Liouville as holding beyond the semiclassical limit.} In fact, there are many reasons why Liouville field theory should not be regarded as dual to a sensible quantum gravity theory: the continuous spectrum and the absence of an $SL(2,\mathbb{C})$-invariant vacuum are probably the most salient reasons. Nevertheless, nothing prevents us from investigating to what extent this relation between AdS$_3$ gravity and Liouville CFT$_2$ can be taken as valid. 

Therefore, let us be reminded of some basic aspects of Liouville field theory. The theory has a central charge given by 
\begin{equation}
c=1+6Q^2 \, , \ \ \text{with}\ \ Q=q+q^{-1},\label{cL} 
\end{equation}
where $q\in \mathbb{R}$ for $c\geq 25$, and $Q$ is the background charge. The semiclassical (large $c$) limit of the theory corresponds to $q\to 0$. 

Liouville field theory has a continuous spectrum, with normalizable states having conformal dimension
\begin{equation}
h=\bar{h}= \frac{c-1}{24}+\lambda ^2 \, , \ \ \text{with}\ \ \lambda^2\in \mathbb{R}_{\geq 0}.\label{arr}
\end{equation}
We notice from (\ref{arr}) that the spectrum has a gap, which in the semiclassical limit reads
\begin{equation}
\text{min}(h)=\text{min}(\bar{h})= \frac{c-1}{24} \simeq \frac{1}{4q^2};\label{fff}
\end{equation}
cf. (\ref{Gfff}). From (\ref{ffff}) and (\ref{fff}) we read a relation between the gravity parameter $\ell/G$ and the Liouville variable $q$; namely
\begin{equation}
q^2=\frac{4G}{\ell}\, .
\end{equation}

In addition to the normalizable states (\ref{arr}), the theory has other interesting lower-weight representations. These are the degenerate representations; namely, non-normalizable, spinless states that contain null descendants in the Verma module and are useful to carry on the bootstrap method in the theory, for example, to solve correlation functions. The states of the degenerate representations are of the form 
\begin{equation}
h_{m,n}=\bar{h}_{m,n}= \frac{c-1}{24}-\frac14 (mq+nq^{-1})^2\, , \ \ \text{with}\ \ m,n\in \mathbb{Z}_{\geq 0}.\label{arr1}
\end{equation}
If we again consider the semiclassical limit, we can write the following state identification
\begin{equation}
h_{m,n}=\bar{h}_{m,n}\simeq  \frac{c}{24}(1-n^2) \simeq  \frac{\ell}{16G}(1-n^2) \,,
\end{equation}
which yields the mass and angular momentum of the BPS states,
\begin{equation}
 M\simeq - \frac{n^2}{8G} \, , \qquad J=0\, .
\end{equation}

This means that we can identify the BPS static configurations of negative mass with the semiclassical limit of Liouville degenerate representations.

Now, let us consider the super-Liouville theory: Virasoro central charge in the superconformal algebra of super-Liouville theory is
\begin{equation}
\hat{c}= 1+2Q^2 \, .
\end{equation}
On the other hand, the central charge $\hat{c}$ is also related to the Chern-Simons level $k$ as $\hat{c}= 4k$. It is useful to compare the notation in \cite{Belavin:2006pv} with that in \cite{Banados:1998pi}. To do so, we can define 
\begin{equation}
\hat c\, =\,\frac 23\, {c},
\end{equation}
with $c=6k=\frac{3\ell}{2G}$ being the Brown-Henneaux central charge. 

Studying the degenerate representations in super-Liouville, we also find
\begin{equation}
h_{m,n}+\bar{h}_{m,n}-\frac{c}{12}=\ell M =-\frac{n^2\ell }{8G}\,.
\end{equation}

\begin{figure}%[H]
    \centering
\includegraphics[width=0.980\textwidth]{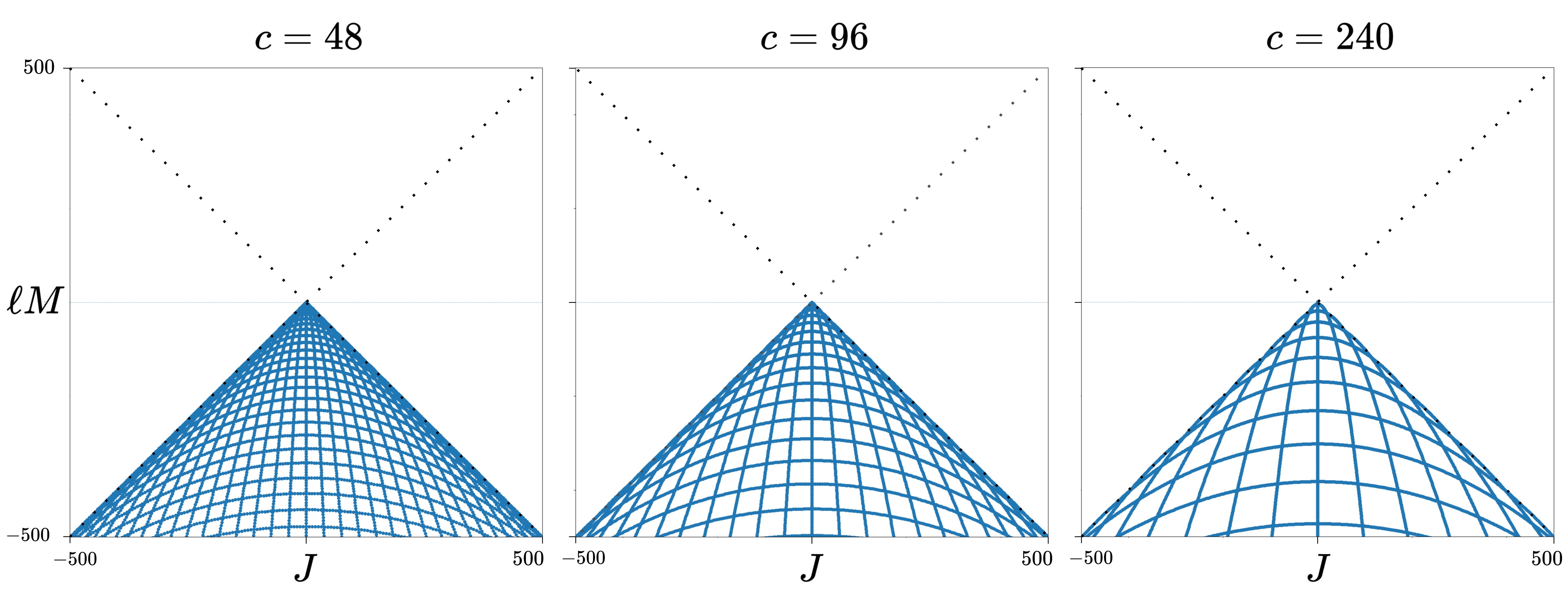}
    \caption{Non-diagonal representations for some specific, finite values of $c$ as a function of $M$ and $J$. A sparse spectrum is observed for large $c$.}
\label{Figura}
\end{figure}

So far, we have discussed spin-zero representations, which correspond to static geometries. For those states, we have identified the BPS configurations of negative mass with the semiclassical limit of spinless Liouville degenerate representations. The question arises whether such an identification is also possible for the spinning ($J\neq 0 $) configurations. Answering this question leads us to investigate the non-diagonal representations of CFT$_2$, which were recently studied in \cite{Migliaccio}. These representations take the form
\begin{equation}
h_{m,n}=\frac{c-1}{24}-\frac{1}{4}(mq+nq^{-1})^2 \, , \ \ \  \bar h _{m,n}=\frac{c-1}{24}-\frac{1}{4}(mq-nq^{-1})^2, \label{DEJKL}
\end{equation}
with $m,n$ taking (semi-)integer values, cf. \cite{Miglacho2}. Figure \ref{Figura} depicts the non-diagonal representations of \cite{Migliaccio} for some (finite) values of the central charge $c$. This yields the spin $h_{m,n}-\bar h_{m,n}=mn$ and, in the semiclassical limit, the condition
\begin{equation}
h_{m,n} -\frac{c}{24}=\frac12 (\ell M + J) \simeq -\frac{n^2\ell}{16G},\label{ultima}
\end{equation}
which is exactly the BPS condition (\ref{global}) after restoring the factor of $G$. However, the last approximation in (\ref{ultima}) requires $h_{m,n}-\bar h_{m,n}$ to be parametrically small relative to $\frac{n^2}{q^2}$ in the $q\to 0$ limit, and so this fails to represent states with non-zero spin in the semiclassical limit. 

It would be interesting to explore the CFT$_2$ realization of spinning BPS states discussed here in the semiclassical limit from a boundary perspective. This would require to take in (\ref{DEJKL}) $m\sim\hat{m}q^{-2}$, which yields $\ell M+J\simeq -\frac{(\hat{m}+n)^2\ell}{8G}$. This result can be thought of as a motivation to further study non-diagonal representations of supersymmetric non-rational CFT$_2$ and generalize the results of \cite{Migliaccio, Miglacho2}. In fact, the representations studied in those works do not suffice to describe the full spectrum of AdS$_3$ gravity solutions in the semiclassical limit. A simple way to see this is by noticing that, for sufficiently large central charge $c$, the non-diagonal representations of \cite{Migliaccio} correspond to non-unitary states, and they cannot represent spinning BTZ black hole geometries. This manifestly shows that a deeper study of such non-diagonal representations in non-compact CFT is needed. The work of Migliaccio and Ribault \cite{Migliaccio} can be thought of as a first step in that direction. It would also be interesting to explore the relation between the solutions studied here and some of the states discussed in Ref.~\cite{Balog:1997zz}. Understanding the connection between theories in two dimensions would also be important to seek. We leave these problems for the future. 

%%%%%%%%%%%%%%%%%%%%%%%%%%%%%%%%%%%%%%%%%%%%%%
\section{Discussion}
%%%%%%%%%%%%%%%%%%%%%%%%%%%%%%%%%%%%%%%%%%%%%%%

In this paper, we surveyed the BTZ family of stationary solutions of 3D AdS gravity, which admit globally defined Killing spinors. These BPS states notoriously include naked singularities such as conical defects and excesses. The solutions were examined by formulating the theory as a Chern-Simons gauge theory on $\mathrm{OSp}(2|1) \times \mathrm{OSp}(2|1)$. The Killing spinor equation were solved on the stationary BTZ geometry looking for solutions on the entire space of parameters. Special attention was given to naked singularities that represent geometries with integer angular excesses corresponding to values of the BTZ parameters, obeying $M\pm J/l=-n^2$, $n\in \mathbb{Z}$. These solutions generalize previous analyses of the parameter space of BTZ geometry in AdS$_3$ supergravity. We argue that, in the semiclassical limit, these BPS states can be associated to degenerate representations of the Virasoro symmetry at the boundary. The degenerate representations to which the BPS states correspond, while not being unitary representations, are somehow special and play an important role in the non-rational CFT at the quantum level. For instance, these are the representations involved in the so-called ``Teschner trick'' when computing the correlation functions with bootstrap techniques. Besides, they appear in the Zamolodchikov higher order differential equations obeyed by correlation functions in Liouville theory. They also appear in the formulae that relate correlation functions of Liouville with those of the non-compact WZW theory. The case of non-diagonal representations, $J\neq 0$, which describe stationary geometries with angular excesses, are less understood; they are associated with Virasoro representations of non-rational CFTs and deserve further study.

It may seem surprising that the sector corresponding to naked singularities contains infinitely many BPS configurations that could represent perturbatively stable vacua. Of course, the fact that metrics with $M\pm|J|/\ell=-n^2$ admit a Killing spinor may not be sufficient to qualify a geometry as a vacuum state, as it may not correspond to a manifold obtained by a genuine identification of AdS, for example. Here we refer to perturbative stability as the consequence of a saturated BPS condition in a supersymmetric system. BPS states could be destabilized if matter fields are introduced as, for instance, in \cite{Casals:2016}, where it is shown that naked singularities can decay into black holes in the presence of a quantum scalar field. A deformation of this nature changes the vacuum structure of the theory, generically shifting the vacuum and introducing new channels for decays to more stable states. It has also been shown that the overspinning geometries are ill-behaved under perturbative corrections \cite{Baake2023}, and it is therefore doubtful that they could define stable vacuum states. In any case, the rich structure that emerges in the sector $M<|J|/\ell$ suggests that perhaps the contribution of those configurations to the partition function should be taken into account, beyond the conical defects \cite{2004.14428}.

The situation for naked singularities obtained by going to the negative mass spectrum of higher-dimensional black holes is radically different because the curvature blows up as $r\to 0$. This is unlike the situation in $2+1$ dimensions, where the curvature remains constant everywhere except in $r=0$. Naked singularities in the negative mass spectrum of higher-dimensional black holes would probably be unstable. Alternatively, higher-dimensional branes obtained by identifications in AdS$_n$, could provide BPS states as in \cite{Edelstein:2010, Edelstein:2011}. 

It is reassuring that the solutions \eqref{rotating} match those in the seminal work of Coussaert and Henneaux \cite{Coussaert:1993jp}, when the mass and angular momentum parameters are restricted to the range $M\geq |J|/\ell$ for the appropriate identifications for $\eta_1$ and $\eta_2$. Our results can be extended to the case in which the geometry also includes torsion \cite{ANTZ}, which suggests that BPS naked singularities are generic in 2+1 AdS geometries. 

%%%%%%%%%%%%%%%%%%%%%%%%%%%%%%%%%%%%%
\subsection*{Acknowledgments}
%%%%%%%%%%%%%%%%%%%%%%%%%%%%%%%%%%%%%
The authors thank Laura Andrianopolis, Glenn Barnich, Laura Donnay, Andrea Campoleoni, Marc Henneaux, Cristian Mart\'inez, Ruggero Noris, Julio Oliva, Joris Raeymaekers, Sylvain Ribault, Mario Trigiante, and Gustavo Turiaci for discussions, correspondence and references. This work has been funded in part by Anillo Grant ANID/ACT210100 and FONDECYT Grants 1220862, 1230112, 1230492, 1231779 and 1241835.


\begin{thebibliography}{9}
%1%
\bibitem{Deser-Zumino} S.~Deser and B.~Zumino, ``Consistent Supergravity,'' Phys. Lett. B \textbf{62} (1976), 335. doi:10.1016/0370-2693(76)90089-7
%2%
\bibitem{DJtH} S.~Deser, R.~Jackiw and G.~'t Hooft, ``Three-Dimensional Einstein Gravity: Dynamics of Flat Space,'' Annals Phys. \textbf{152} (1984), 220.
doi:10.1016/0003-4916(84)90085-X
%3%
\bibitem{BTZ}
M.~Ba\~nados, C.~Teitelboim and J.~Zanelli,
``The Black hole in three-dimensional space-time,''
Phys. Rev. Lett. \textbf{69} (1992), 1849-1851.[arXiv:hep-th/9204099 [hep-th]]. doi:10.1103/PhysRevLett.69.1849
%4%
\bibitem{BH}
J.~D.~Brown and M.~Henneaux, ``Central Charges in the Canonical Realization of Asymptotic Symmetries: An Example from Three-Dimensional Gravity,'' Commun. Math. Phys. \textbf{104} (1986), 207-226. doi:10.1007/BF01211590
%5%
\bibitem{BHTZ}
M.~Ba\~nados, M.~Henneaux, C.~Teitelboim and J.~Zanelli, ``Geometry of the (2+1) black hole,''
Phys. Rev. D \textbf{48} (1993), 1506-1525. [erratum: Phys. Rev. D \textbf{88} (2013), 069902.]
[arXiv:gr-qc/9302012 [gr-qc]]. doi:10.1103/PhysRevD.48.1506
%6%
\bibitem{Olivera-Z}
O.~Mi\v{s}kovi\'c and J.~Zanelli, ``On the negative spectrum of the 2+1 black hole,'' Phys. Rev. D \textbf{79} (2009), 105011. [arXiv:0904.0475 [hep-th]].
doi:10.1103/PhysRevD.79.105011
%7%
\bibitem{Briceno-Martinez-Z} M. Brice\~no, C. Mart\'inez and J. Zanelli, ``Central Singularity of the BTZ Geometries,'' Phys. Rev. D \textbf{110} (2024) no.2, 024075. [arXiv:2404.06552 [gr-qc]]. doi:10.1103/PhysRevD.110.024075
%8%
\bibitem{Witten}
E.~Witten, ``Three-Dimensional Gravity Revisited,''
[arXiv:0706.3359 [hep-th]].
%9%
\bibitem{Strominger}
A.~Strominger, ``Black hole entropy from near horizon microstates,'' JHEP \textbf{02} (1998), 009. [arXiv:hep-th/9712251 [hep-th]]. doi:10.1088/1126-6708/1998/02/009
%10%
\bibitem{Coussaert:1993jp}
O.~Coussaert and M.~Henneaux, ``Supersymmetry of the (2+1) black holes,'' Phys. Rev. Lett. \textbf{72} (1994), 183-186. [arXiv:hep-th/9310194 [hep-th]]. doi:10.1103/PhysRevLett.72.183
%11%
\bibitem{CG}
J.~Oliva, D.~Tempo and R.~Troncoso,
``Static spherically symmetric solutions for conformal gravity in three dimensions,'' Int. J. Mod. Phys. A \textbf{24} (2009), 1588-1592. [arXiv:0905.1510 [hep-th]].
doi:10.1142/S0217751X09045054
%12%
\bibitem{Garcia}
A.~A.~Garc\'ia, F.~W.~Hehl, C.~Heinicke and A.~Macias,
``Exact vacuum solution of a (1+2)-dimensional Poincare gauge theory: BTZ solution with torsion,'' Phys. Rev. D \textbf{67} (2003), 124016. [arXiv:gr-qc/0302097 [gr-qc]]. doi:10.1103/PhysRevD.67.124016
%13%
\bibitem{HorowitzWelch}
G.~T.~Horowitz and D.~L.~Welch, ``Exact three-dimensional black holes in string theory,'' Phys. Rev. Lett. \textbf{71} (1993), 328-331. [arXiv:hep-th/9302126 [hep-th]]. doi:10.1103/PhysRevLett.71.328
%14%
\bibitem{TMG}
K.~A.~Moussa, G.~Clement and C.~Leygnac, ``The Black holes of topologically massive gravity,'' Class. Quant. Grav. \textbf{20} (2003), L277-L283. [arXiv:gr-qc/0303042 [gr-qc]]. doi:10.1088/0264-9381/20/24/L01
%15%
\bibitem{NMG}
E.~A.~Bergshoeff, O.~Hohm and P.~K.~Townsend, ``More on Massive 3D Gravity,'' Phys. Rev. D \textbf{79} (2009), 124042. [arXiv:0905.1259 [hep-th]]. doi:10.1103/PhysRevD.79.124042
%16%
\bibitem{Banados}
M.~Ba\~nados and S.~Theisen, ``Three-dimensional massive gravity and the bigravity black hole,'' JHEP \textbf{11} (2009), 033. [arXiv:0909.1163 [hep-th]]. doi:10.1088/1126-6708/2009/11/033
%17%
\bibitem{Krauss}
M.~Gutperle and P.~Kraus, ``Higher Spin Black Holes,'' JHEP \textbf{05} (2011), 022. [arXiv:1103.4304 [hep-th]]. doi:10.1007/JHEP05(2011)022
%18%
\bibitem{1106.4788}
M.~Ammon, M.~Gutperle, P.~Kraus and E.~Perlmutter, ``Spacetime Geometry in Higher Spin Gravity,'' JHEP \textbf{10} (2011), 053. [arXiv:1106.4788 [hep-th]]. doi:10.1007/JHEP10(2011)053
%19%
\bibitem{1108.2567}
P.~Kraus and E.~Perlmutter, ``Partition functions of higher spin black holes and their CFT duals,'' JHEP \textbf{11} (2011), 061. [arXiv:1108.2567 [hep-th]]. doi:10.1007/JHEP11(2011)061
%20%
\bibitem{1203.0015}
M.~R.~Gaberdiel, T.~Hartman and K.~Jin, ``Higher Spin Black Holes from CFT,'' JHEP \textbf{04} (2012), 103.
[arXiv:1203.0015 [hep-th]]. doi:10.1007/JHEP04(2012)103
%21%
\bibitem{1302.0816}
J.~de Boer and J.~I.~Jottar, ``Thermodynamics of higher spin black holes in $AdS_3$,'' JHEP \textbf{01} (2014), 023. [arXiv:1302.0816 [hep-th]]. doi:10.1007/JHEP01(2014)023
%22%
\bibitem{1309.4362}
M.~Henneaux, A.~P\'erez, D.~Tempo and R.~Troncoso, ``Chemical potentials in three-dimensional higher spin anti-de Sitter gravity,'' JHEP \textbf{12} (2013), 048. [arXiv:1309.4362 [hep-th]]. doi:10.1007/JHEP12(2013)048
%23%
\bibitem{1402.1465}
A.~P\'erez, D.~Tempo and R.~Troncoso, ``Higher Spin Black Holes,'' Lect. Notes Phys. \textbf{892} (2015), 265-288.
[arXiv:1402.1465 [hep-th]]. doi:10.1007/978-3-319-10070-8\_10
%24%
\bibitem{1301.0847}
A.~P\'erez, D.~Tempo and R.~Troncoso,
``Higher spin black hole entropy in three dimensions,''
JHEP \textbf{04} (2013), 143. [arXiv:1301.0847 [hep-th]]. doi:10.1007/JHEP04(2013)143
%25%
\bibitem{1207.2844}
A.~P\'erez, D.~Tempo and R.~Troncoso, ``Higher spin gravity in 3D: Black holes, global charges and thermodynamics,'' Phys. Lett. B \textbf{726} (2013), 444-449. [arXiv:1207.2844 [hep-th]]. doi:10.1016/j.physletb.2013.08.038
%26%
\bibitem{1512.00073}
M.~Ba\~nados, A.~Castro, A.~Faraggi and J.~I.~Jottar, ``Extremal Higher Spin Black Holes,'' JHEP \textbf{04} (2016), 077. [arXiv:1512.00073 [hep-th]]. doi:10.1007/JHEP04(2016)077
%27%
\bibitem{1111.3381}
A.~Castro, R.~Gopakumar, M.~Gutperle and J.~Raeymaekers, ``Conical Defects in Higher Spin Theories,'' JHEP \textbf{02} (2012), 096.[arXiv:1111.3381 [hep-th]]. doi:10.1007/JHEP02(2012)096
%28%
\bibitem{Edery:2020kof}
A.~Edery, ``Non-singular vortices with positive mass in 2+1 dimensional Einstein gravity with AdS$_3$ and Minkowski background,'' JHEP \textbf{01} (2021), 166.
[arXiv:2004.09295 [hep-th]]. doi:10.1007/JHEP01(2021)166
%29%
\bibitem{Edery:2022crs}
A.~Edery, ``Nonminimally coupled gravitating vortex: Phase transition at critical coupling \ensuremath{\xi}c in AdS3,'' Phys. Rev. D \textbf{106} (2022) no.6, 065017. [arXiv:2205.12175 [hep-th]]. doi:10.1103/PhysRevD.106.065017
%30%
\bibitem{Migliaccio}
S.~Migliaccio and S.~Ribault, ``The analytic bootstrap equations of non-diagonal two-dimensional CFT,'' JHEP \textbf{05} (2018), 169. [arXiv:1711.08916 [hep-th]]. doi:10.1007/JHEP05(2018)169
%31%
\bibitem{1210.8452}
E.~Perlmutter, T.~Prochazka and J.~Raeymaekers, ``The semiclassical limit of $W_N$ CFTs and Vasiliev theory,'' JHEP \textbf{05} (2013), 007. [arXiv:1210.8452 [hep-th]]. doi:10.1007/JHEP05(2013)007
%32%
\bibitem{1712.08078}
A.~Campoleoni, S.~Fredenhagen and J.~Raeymaekers, ``Quantizing higher-spin gravity in free-field variables,'' JHEP \textbf{02} (2018), 126. [arXiv:1712.08078 [hep-th]]. doi:10.1007/JHEP02(2018)126
%33%
\bibitem{1412.0278}
J.~Raeymaekers, ``Quantization of conical spaces in 3D gravity,'' JHEP \textbf{03} (2015), 060. [arXiv:1412.0278 [hep-th]]. doi:10.1007/JHEP03(2015)060
$34$
\bibitem{2012.07934}
J.~Raeymaekers, ``Conical spaces, modular invariance and $c_{p,1}$ holography,'' JHEP \textbf{03} (2021), 189. [arXiv:2012.07934 [hep-th]]. doi:10.1007/JHEP03(2021)189
%35%
\bibitem{1811.05710}
S.~Collier, Y.~Gobeil, H.~Maxfield and E.~Perlmutter,
``Quantum Regge Trajectories and the Virasoro Analytic Bootstrap,'' JHEP \textbf{05} (2019), 212. [arXiv:1811.05710 [hep-th]]. doi:10.1007/JHEP05(2019)212
%36%
\bibitem{2004.14428}
N.~Benjamin, S.~Collier and A.~Maloney, ``Pure Gravity and Conical Defects,'' JHEP \textbf{09} (2020), 034. [arXiv:2004.14428 [hep-th]]. doi:10.1007/JHEP09(2020)034
%37%
\bibitem{1904.05228}
T.~G.~Mertens and G.~J.~Turiaci, ``Defects in Jackiw-Teitelboim Quantum Gravity,'' JHEP \textbf{08} (2019), 127. [arXiv:1904.05228 [hep-th]]. doi:10.1007/JHEP08(2019)127
%38%
\bibitem{2011.01953}
M.~Heydeman, L.~V.~Iliesiu, G.~J.~Turiaci and W.~Zhao, ``The statistical mechanics of near-BPS black holes,'' J. Phys. A \textbf{55}, no.1 (2022), 014004. [arXiv:2011.01953 [hep-th]]. doi:10.1088/1751-8121/ac3be9
%39%
\bibitem{2305.19438}
G.~J.~Turiaci and E.~Witten, ``$\mathcal{N}=2$ JT Supergravity and Matrix Models,'' JHEP \textbf{12} (2023), 003. [arXiv:2305.19438 [hep-th]]. doi:10.1007/JHEP12(2023)003
%40%
\bibitem{Romans}
L.~J.~Romans, ``Black holes in cosmological Einstein-Maxwell theory,'' Workshop on Superstrings and Related Topics, 416-427 (1992).
%41%
\bibitem{Achucarro:1986uwr}
A.~Ach\'ucarro and P.~K.~Townsend, ``A Chern-Simons Action for Three-Dimensional anti-De Sitter Supergravity Theories,'' Phys. Lett. B \textbf{180} (1986), 89. doi:10.1016/0370-2693(86)90140-1
%42%
\bibitem{Miskovic:2006tm} O.~Mi\v{s}kovi\'c and R.~Olea, ``On boundary conditions in three-dimensional AdS gravity,'' Phys. Lett. B \textbf{640} (2006), 101-107. [arXiv:hep-th/0603092 [hep-th]]. doi:10.1016/j.physletb.2006.07.045
%43%
\bibitem{Izquierdo:1994jz}
J.~M.~Izquierdo and P.~K.~Townsend, ``Supersymmetric space-times in (2+1) adS supergravity models,'' Class.~Quant.~Grav.~\textbf{12} (1995), 895-924. [arXiv:gr-qc/9501018 [gr-qc]]. doi:10.1088/0264-9381/12/4/003
%44%
\bibitem{Balasubramanian:2000rt}
V.~Balasubramanian, J.~de Boer, E.~Keski-Vakkuri, and S.~Ross, ``Supersymmetric conical defects: Towards a string theoretic description of black hole formation,'' Phys.~Rev.~D \textbf{64} (2001), 064011. [arXiv:hep-th/0011217 [hep-th]]. doi:10.1103/PhysRevD.64.064011
%45%
\bibitem{Donnay}
G.~Barnich, L.~Donnay, J.~Matulich and R.~Troncoso, ``Asymptotic symmetries and dynamics of three-dimensional flat supergravity,'' JHEP \textbf{08} (2014), 071. [arXiv:1407.4275 [hep-th]]. doi:10.1007/JHEP08(2014)071
%46%
\bibitem{vanDriel}
O.~Coussaert, M.~Henneaux and P.~van Driel, ``The Asymptotic dynamics of three-dimensional Einstein gravity with a negative cosmological constant,'' Class. Quant. Grav. \textbf{12} (1995), 2961-2966. [arXiv:gr-qc/9506019 [gr-qc]]. doi:10.1088/0264-9381/12/12/012
%47%
\bibitem{Krasnov}
K.~Krasnov, ``3-D gravity, point particles and Liouville theory,'' Class. Quant. Grav. \textbf{18} (2001), 1291-1304. [arXiv:hep-th/0008253 [hep-th]]. doi:10.1088/0264-9381/18/7/311
%48%
\bibitem{Donnay2}
L.~Donnay, ``Asymptotic dynamics of three-dimensional gravity,'' PoS \textbf{Modave2015} (2016), 001. [arXiv:1602.09021 [hep-th]]. doi:10.22323/1.271.0001
%49%
\bibitem{Hikida:2012} Y.~Hikida, ``Conical defects and $N=2$ higher spin holography,'' JHEP \textbf{08} (2013), 127. [arXiv:1212.4124 [hep-th]]. doi:10.1007/JHEP08(2013)127
%50%
\bibitem{Giacomini:2006dr}
A.~Giacomini, R.~Troncoso and S.~Willison, ``Three-dimensional supergravity reloaded,'' Class. Quant. Grav. \textbf{24} (2007), 2845-2860. [arXiv:hep-th/0610077 [hep-th]]. doi:10.1088/0264-9381/24/11/005
%51%
\bibitem{Banados:1998pi}
M.~Ba\~nados, K.~Bautier, O.~Coussaert, M.~Henneaux and M.~Ortiz, ``Anti-de Sitter / CFT correspondence in three-dimensional supergravity,'' Phys. Rev. D \textbf{58} (1998), 085020. [arXiv:hep-th/9805165 [hep-th]]. doi:10.1103/PhysRevD.58.085020
%52%
\bibitem{Malda}
J.~M.~Maldacena, ``The Large N limit of superconformal field theories and supergravity,'' Adv. Theor. Math. Phys. \textbf{2} (1998), 231-252. [arXiv:hep-th/9711200 [hep-th]]. doi:10.4310/ATMP.1998.v2.n2.a1
%53%
\bibitem{Belavin:2006pv}
A.~Belavin and A.~Zamolodchikov, ``Higher equations of motion in N = 1 SUSY Liouville field theory,'' JETP Lett. \textbf{84} (2006), 418-424. [arXiv:hep-th/0610316 [hep-th]]. doi:10.1134/S0021364006200033
%54%
\bibitem{Miglacho2}
S.~Migliaccio, ``Conformal bootstrap in two-dimensional conformal field theories with non-diagonal spectrums,'' [arXiv:1901.10922 [hep-th]].
%55%
\bibitem{Balog:1997zz}
J.~Balog, L.~Feher and L.~Palla, ``Coadjoint orbits of the Virasoro algebra and the global Liouville equation,'' Int. J. Mod. Phys. A \textbf{13} (1998), 315-362. [arXiv:hep-th/9703045 [hep-th]]. doi:10.1142/S0217751X98000147
%56%
\bibitem{Casals:2016} M.~Casals, A.~Fabbri, C.~Mart\'\i{}nez, and J.~Zanelli, ``Quantum Backreaction on Three-Dimensional Black Holes and Naked Singularities,'' Phys. Rev. Lett. \textbf{118} (2017), 131102. [arXiv:1608.05366[gr-qc]]. doi:10.1103/PhysRevLett.118.131102
%57%
\bibitem{Baake2023} O.~Baake and J.~Zanelli, ``Quantum backreaction for overspinning BTZ geometries,'' Phys. Rev. D \textbf{107} (2023) no.8, 084015. [arXiv:2301.04256 [hep-th]]. doi:10.1103/PhysRevD.107.084015
%58%
\bibitem{Edelstein:2010} J.~D.~Edelstein, A.~Garbarz, O.~Miskovic, and J.~Zanelli, ``Stable p-branes in Chern-Simons AdS supergravities,'' Phys. Rev. D, \textbf{82} (2010), 044053. [arXiv:1006.3753 [hep-th]]. doi:10.1103/PhysRevD.82.044053
%59%
\bibitem{Edelstein:2011} J.~D.~Edelstein, A.~Garbarz, O.~Miskovic, and J.~Zanelli, ``Geometry and stability of spinning branes in AdS gravity,'' Phys. Rev. D, \textbf{84} (2011), 104046.[arXiv:1108.3523[hep-th]]. doi:10.1103/PhysRevD.84.104046
%60%
\bibitem{ANTZ} L.~Andrianopoli, R.~Noris, M.~Trigiante and J.~Zanelli, ``Supersymmetric States in Anti-de Sitter D=3 Supergravity with Chiral Torsion,'' Phys. Rev. Lett. \textbf{133}, no.3 (2024), 031602. [arXiv:2404.12427 [hep-th]]. doi:10.1103/PhysRevLett.133.031602

\end{thebibliography}
\end{document}